\documentclass{ws-procs9x6}

\usepackage{graphicx}
\newcommand{\be}{\begin{eqnarray}}
\newcommand{\ee}{\end{eqnarray}}

\begin{document}

\title{Deconfinement, Monopoles \\
and New Phenomena in Heavy 
Ion Collisions}

\author{Edward Shuryak}

\address{ Department of Physics and Astronomy, University at Stony Brook,\\
Stony Brook NY 11794 USA\\
$^*$E-mail: shuryak@tonic.physics.sunysb.edu}

\begin{abstract}
We discuss various manifestations of the ``magnetic scenario" for the quark-gluon plasma viewed 
as a mixture of two plasmas, of electrically
(quark and gluons) as well as magnetically charged quasiparticles.
Near
the deconfinement phase transition, $T\approx T_c$
very small density of free quarks should lead to negligible screening of
electric field while magnetic screening remains strong.  The consequence of this should be existence of a ``corona"
of the QGP, in a way similar to that of the Sun, in which electric fields influence 
propagation of perturbations and even form metastable flux tubes.
The natural tool for its description is (dual) magnetohydrodynamics: among observable consequences
is splitting  of sound into  two modes,
 with larger and  smaller velocity. The latter can be zero, hinting for formation of pressure-stabilized flux tubes. Remarkably, recent experimental discoveries at RHIC
show effects similar to expected  for  ``corona structures".  In dihadron correlation function with large-$p_t$ trigger there are
 a ``cone'' and a ``hard ridge'',while the so called ``soft ridge'' is a similar structure seen without hard trigger. They seem to be remnants of flux tubes, which -- 
contrary to naive expectations --  seem to break less often in near-$T_c$ matter than do confining strings in vacuum.
\end{abstract}
\vskip .3cm

\bodymatter

\section{Introduction}
   Let me start my written version of the talk in the same way as the oral one:
   with comments on my interaction with Misha Shifman. It is more than 30 years
   since  he start visiting Novosibirsk, working with Arkady on QCD sum rules,
   while I occasionally managed to visit ITEP.
   1970's were formative years for all of us, and I learned a lot from those
   discussions. 
    While we shared
   many interests over these years -- to QCD correlators, heavy quark hadrons, instantons, now monopoles and confinement   -- we have so different styles that we never actually 
   worked on exactly the same thing. That produced some distance, a different perspective and often
   mutual criticism. And yet, it only strengthened our friendship: we are in agreement on so many other things in life. So, it is  a delight
   to see Misha at another conference, of his beloved  ``Continuos Advances" series in particular,  
   full of life and as usual having a joke or a couple of good (if  sceptical) questions to any talk. 
    
This talk fits into this long-time pattern  perfectly. While in the years 2000-2005 I was preoccupied by RHIC data
 and then various strongly coupled plasmas, AdS/CFT and other issues not close to Misha,  my talk at  Continuos Advances 06 included 
a part about ``magnetic scenario" for temperatures near $T_c$, on a plasma side. Presently both of us   
study different aspects of the ``magnetic side" of QCD, trying to understand
 confinement mechanism:  while Misha looks at models with enough supersymmetry to have rigid theoretical control, I looks at RHIC and lattice data plus some
generic theoretical arguments to get the picture. Recent results and experimental discoveries look quite exciting:
 I hope that makes a good addition to this  proceeding, a kind of our collective birthday present. 
%So, as before, we are quite complementary...   

\section{Dual electric/magnetic plasmas and transport}
Let me thus start by reminding what I was speaking about at CA06. Among very few things we can agree on with Misha
is a general  ``dual superconductor" view of  confinement by 't Hooft and Mandelsham. If so,
 the confinement transition must be Bose-Einstein condensation (BEC)
of certain magnetically charged objects. Furthermore,  right above $T_c$ there should be sufficient density of them in a ``normal" (uncondensed) state.
%Thus came the view of dual plasmas, with both electric and magnetic quasiparticles: a system interesting in its own right. Indeed, even in classical approximations
 %it was
%never studied before:  we dont have electromagnetic Dirac monopoles to warrant such studies. As I will describe in the next section, during the years 07-08 
%many of the things we proposed -- such as that electric and magnetic coupling in our two plasmas  {\em do run in the opposite directions} -- were confirmed on the lattice.
%Logically following from this magnetic scenario is a new view \cite{Shuryak:2009cy} on QGP  produced in heavy ion collision, predicting it
%to be surrounded by  a (dual) ``corona", not too dissimilar to that of the Sun. And, most unexpectedly, there are experimental observations 
%which point to unusually propagating perturbations and surprisingly well preserved flux tubes, see the last section.

  Dirac's quantization condition demands the product
of electric and magnetic couplings to be an integer: thus those two couplings {\em must run in the opposite directions} (as it happens e.g. in Seiberg-Witten 
$\cal N$=2 theory). So Liao and myself \cite{Liao_ES_mono} proposed a view of CD plasma as of dual mixture, with two Coulomb-interacting subsystems, both at
 strong coupling whose magnitude is running with $T$ in the $opposite$ directions. (Chernodub and Zakharov \cite{Chernodub:2006gu} have independently 
 proposed a similar picture, with large role of monopoles in the near-$T_c$ region.)
 But how to see if this is true?

\begin{figure}[t]
   \includegraphics[width=5.cm]{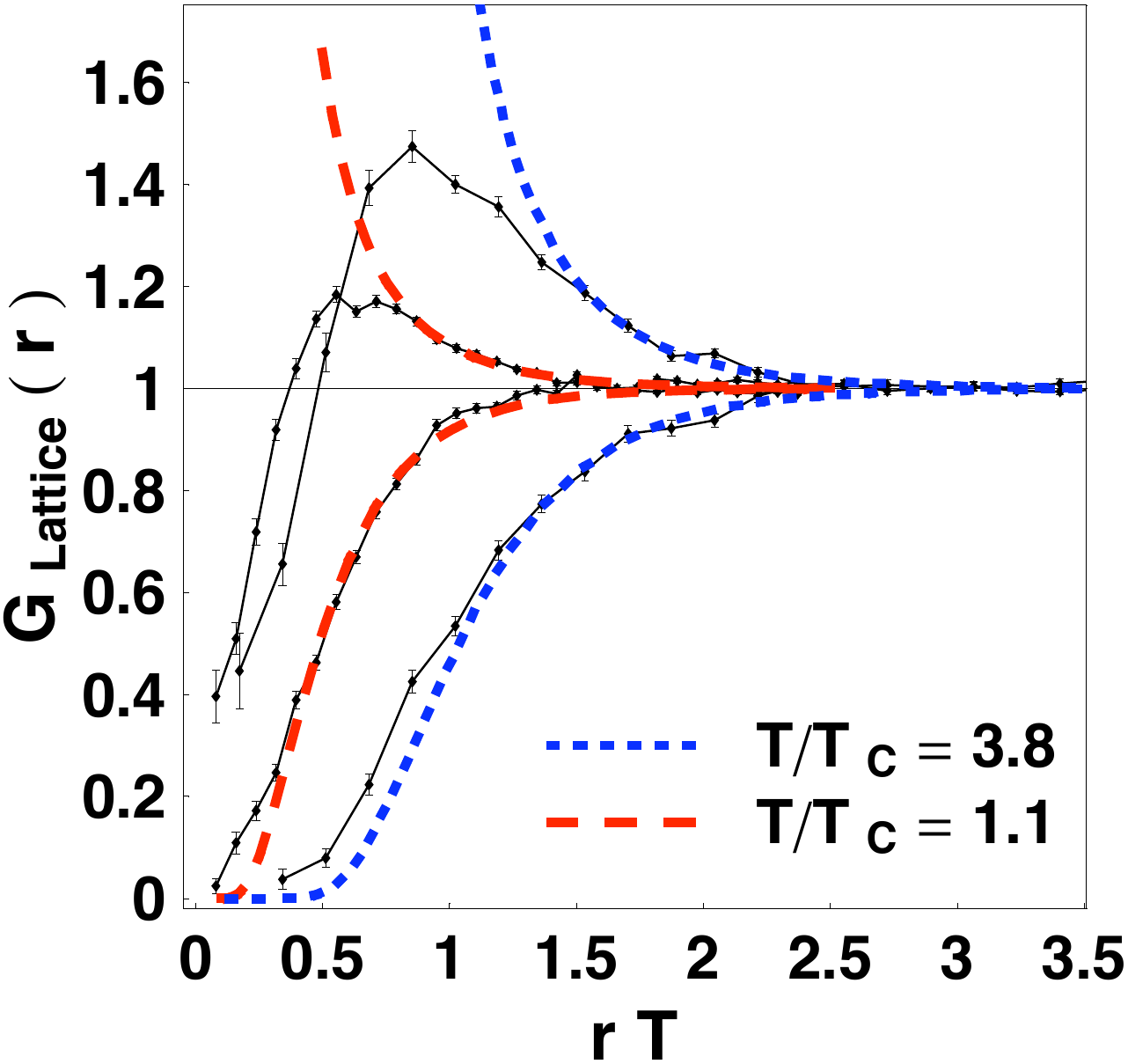}  \includegraphics[width=5.cm]{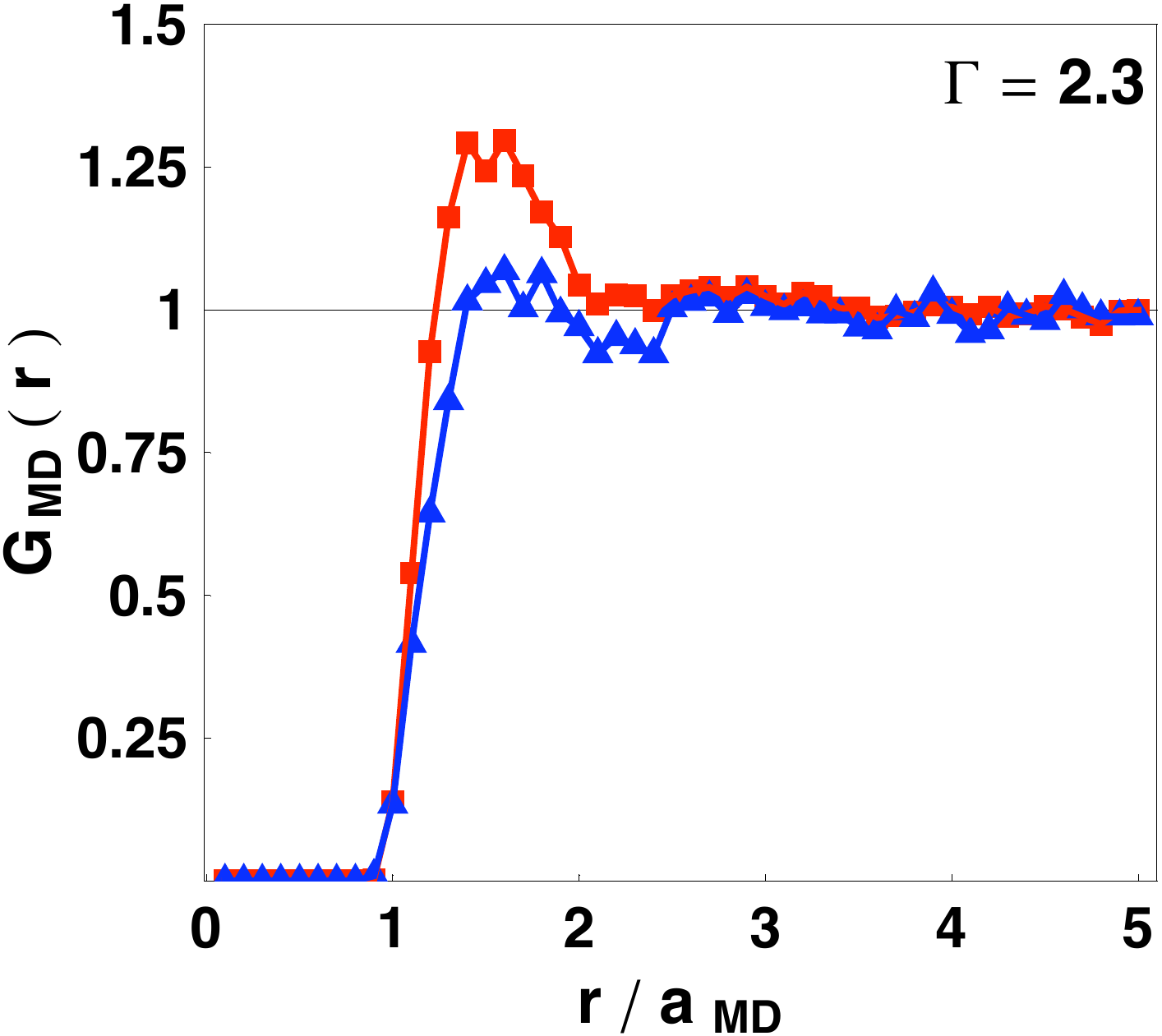}
  \vspace{0.1in}
\caption{ (left) Monopole-antimonopole (the upper two
curves) and monopole-monopole (the lower two curves) correlators
at $1.1T_c$(red long dashed) and $3.8T_c$(blue dot dashed): points
with error bars are lattice data
\protect\cite{D'Alessandro:2007su}, the dashed lines are our fits. 
(right) Monopole-antimonopole (boxes with red curve) and monopole-monopole (triangles with blue curve)
 correlators from MD simulations at $\Gamma=2.3$.
\label{fig_corr} }
\end{figure}

On the lattice one can identify monopoles by ``the end of the Dirac string" in a certain gauge\footnote{Let me note that thus what is called a monopole actually include dyons,
and that small size of the core and coarse lattices does not allowed for detail knowledge of its structure.}:
so one can look at monopole-antimonopole correlations in space. An example 
\cite{D'Alessandro:2007su} for two temperatures is shown in Fig.\ref{fig_corr} (left).
 Note characteristic peaks
  which  are found in strongly correlated liquids.  Note also that
  the higher $T$ has stronger correlation: this implies that
  magnetic Coulomb coupling $does$ grow with $T$, contrary to
  asymptotic freedom for electric coupling. The  right figure is our molecular dynamics (MD)
simulations for plasmas containing monopoles: note that the red one (monopole-antimonopole) has the same shape and magnitude as
the lattice data on the left.  So, a picture  two dual plasmas, electric and magnetic ones, 
does work. 

\begin{figure}[t]
%\begin{minipage}{.48\textwidth}
%\parbox{6cm}{
%\scalebox{.6}{
%\includegraphics{pscatteringrate}\\}}
%\end{minipage}
%\hspace{.4cm}
%\begin{minipage}{.48\textwidth}
%\parbox{6cm}{
%\scalebox{.6}{
\includegraphics[width=4.8cm]{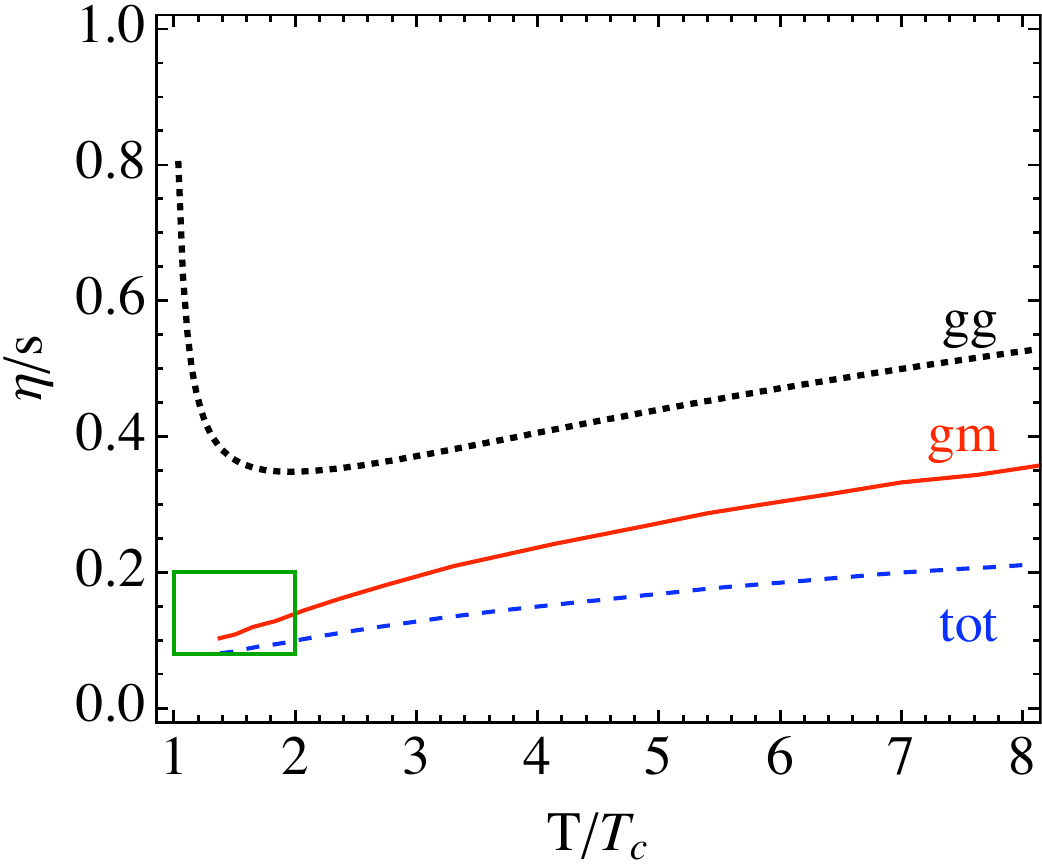} %\\}}
	\includegraphics[width=6cm]{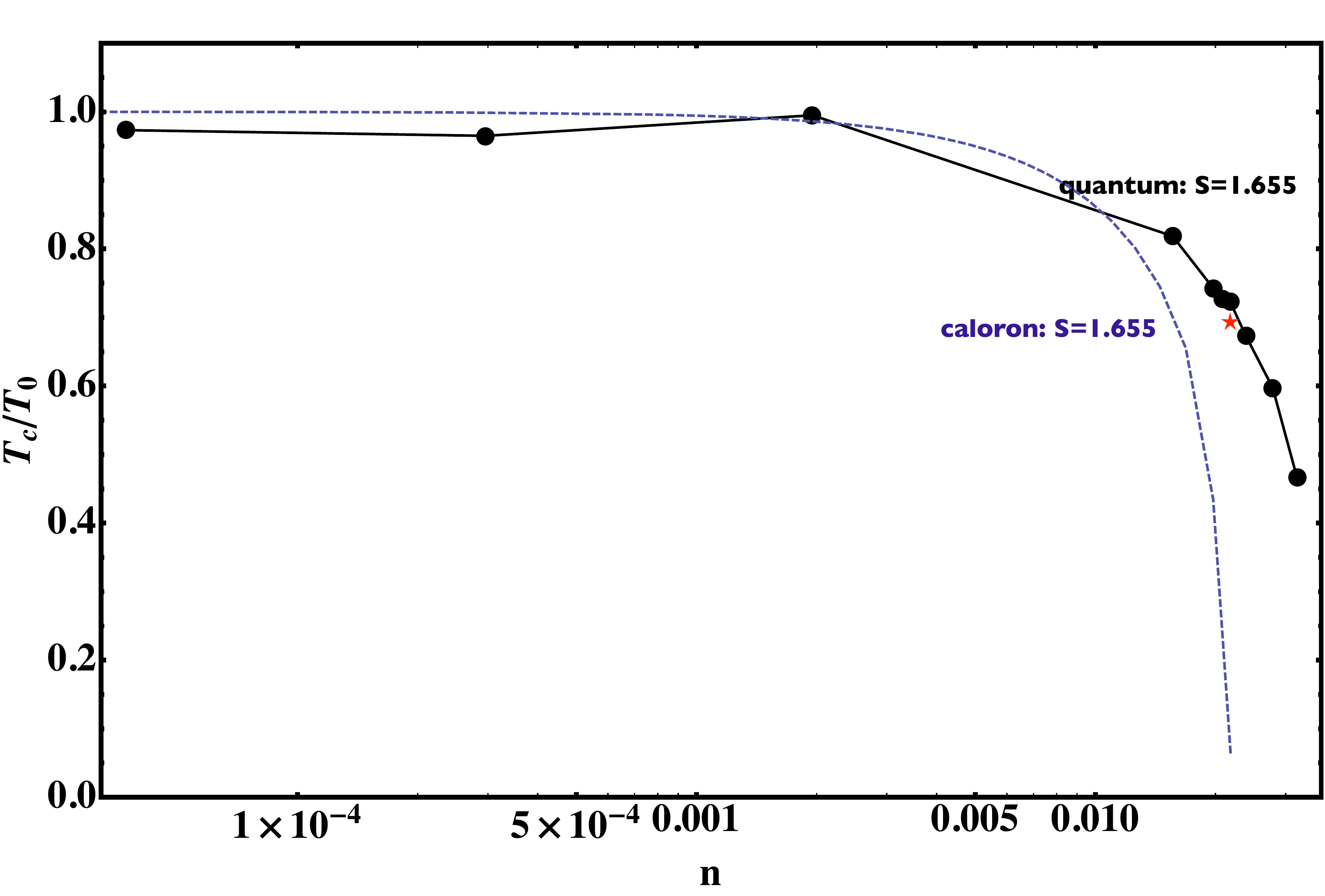}
%\vspace{0.75cm}
\caption[h]{ \label{fig_viscosity} 
%(a) The path for the interacting system and for free one (straight line).
(left) The contributions of the gluon-monopole and gluon-gluon scatterings into viscosity-to-entropy ratio.
(right)
Density dependence of the critical temperature for liquid ${}^4$He: black points are our calculation while the red star is the experimental zero-$T$ critical point.  }
%\end{minipage}
%%\caption{%Left panel: gluon-monopole and gluon-gluon scattering rate. Right panel:
 \end{figure}

      What is the contribution of the monopoles to transport? Before we used  classical molecular dynamics  \cite{Liao_ES_mono}
   while this time I report
     results  by Ratti and myself  \cite{Ratti:2008jz}. We solved  a beautiful quantum-mechanical problem of gluon-monopole scattering.
     Asymptotically, at $T\rightarrow \infty$, this process is sublieading by one power of $log(T)$. Yet
    at $T$ few times $T_c$, when the monopole density is still much smaller
      than that of quarks and gluons, we found that gluon-monopole scattering actually dominates the transport.
     This happens because of characteristic large backward scattering, absent in charge-charge case.  
Scattering rate and  $(\eta/ s)$ are shown in Fig. \ref{fig_viscosity}(left), together with the
ones obtained for the usual perturbative gluon-gluon scattering.
There is a qualitative agreement between these results and the experimental
value for $\eta/s$ observed at RHIC (the box on the left), as well as with the famous AdS/CFT 
result $\eta/s=1/4\pi$.  LHC will create (in PbPb collisions) QGP with $T\sim 4 T_c$: we thus have our prediction
of how good a liquid QGP will be there.

\section{Universal Feynman's criterion for Bose-Einstein condensation and monopoles at deconfinement}
If one wanders  whether confinement is indeed a BEC of monopoles, 
looking at their behavior on the lattice  close to $T_c$  is a good idea. But before doing it,
one should know what to look at.
 In a paper with Marco 
Cristoforetti \cite{Cristoforetti:2009tx} we had revived an old idea by Feynman \cite{fey12}, in which he introduced the notion of the $universal$ critical value for the per-particle action 
in large supercurrent clusters. Note that in Matsubara formalism bosons should make periodic
paths: but it may be not for each particle separately but for say k particles, called k-cluster.
Feynman argued that the sum over cluster size k should diverge at the critical point, when
action suppressing ``jumps" from one particle location to the next is balanced by
multiplicity of clusters. Feynman  included kinetic energy only and  thus approximated it by the straight line, so his expression for the amplitude is %(see Fig.
%stopped here
\be\label{eq:yf}
	y_F=\exp\Big[-\frac{m^*Td^2}{2\hbar^2}\Big]
\ee
%which corresponds to  straight line path, or account for kinetic energy only.

Marco and I decided to work it out more, and include also potential energy.
We had built the ``moving string'' model for the supercurrent, a line of particles
in a crystal (a reasonably good model for liquid He$^4$) and calculated the action
of all the string moving as a whole using realistic atomic potentials. Semiclassical
calculation (another use of  periodic instantons or calorons) or numerical path
integration have given close actions. In Fig.\ref{fig_viscosity}(right) we plot the ratio
of the calculated critical temperature $T_c/T_0$, where $T_0$ is Einstein's critical temperature for  free Bose gas.
We found that for strongly interacting liquid He the critical action is
 $S_c=1.655$, which is {\em exactly the same} value as one would get for noninteracting gas.
So, Feynman was right about universality of his criterion: he just have not done
a complete calculation to verify it for liquid He!

%\begin{figure}
%\hspace{10cm}
%	\includegraphics[width=4cm]{path.pdf}
%	\includegraphics[width=7cm]{tctosin.pdf}
%\vspace{0.75cm}
%\caption[h]{\label{fig_viscosity} 
%(a) The path for the interacting system and for free one (straight line).
%(b)
%Density dependence of the critical temperature. Black points are our calculation,oThe red star is the physical result for ${}^4$He. }
%\end{figure}

Now, with confidence about universal BEC criterion, we applied it for QCD monopoles.
Since we know approximately their density, we can set upper limits on the monopole mass
and magnetic Coulomb coupling at $T_c$. It turns out they must be as light as about 200 MeV or so.

Does it all agree with the lattice behavior of monopoles? As we were told by D'Elia,
clusters with k=1,2,3 indeed have equal probability at $T_c$.
 The monopole mass is not yet accurately measured, but it indeed strongly decreases toward
 $T_c$
about as low as predicted.   In summary:  monopoles do behave similarly to He atoms at
their respective BEC transition temperatures.

%\section{Dual magnetohydrodynamics and flux tubes at and above $T_c$}
\section{QGP corona:  theory }

% shown by squares
%in Fig.\ref{fig_density}.  (The green diamonds on the right-hand
 %side of the plot 
%correspond to the monopole density directly detected on the lattice
 % \citeD'Alessandro:2007su.) Since
% $F=V-TS$ and it has  no linear potential near and above
 %$T_c$, the energy and etropy should cancel each other.
% The summary of those studies is that metastable flux tubes 
%seem to exist   
%at   $T<1.4T_c$, changing from stable to metastable around $T_c$.

The word ``corona" used here comes from the physics of the Sun. Let me briefly remind  that 
it was started by Galileo Galilei, who in 1612 spent some time observing the motion of the black spots on the Sun and correctly concluded from  motion
 of the spots that they must resign on a surface of a rotating sphere: he thus argued the spots were not 
 shadows of some planets passing in front, as it was thought of before. In due time relation between the spots and solar magnetism
 was understood: modern telescopes allows one to see the
fine structure of solar spots, resolving  individual magnetic flux tubes. 
Better understanding of solar magnetism came with the advance of plasma physics in 1940's and development of MHD, which explained both the influence of
diffuse magnetic field on plasma and formation and 
 mechanical stability of the flux tubes. 
%The MHD flux tubes are supported by the electron current, while the positive charges -- the ions -- are heavy and dont move. Since it is not a supercurrent,
%there is inevitable friction and thus metastability of the flux tube solutions. 

Early stages of heavy ion collisions are believed to be described by the
so called ``glasma" ,  a set of random color fields 
created by color charges of partons of  the two colliding nuclei
at the moment of the collision. 
%As argued by McLerran and Venugopalan \cite{McLerran:1993ni}, the 
%color charges of partons at particular location in the transverse
%plane cannot be not correlated (they come from different nucleons) ,
 % thus they add up randomly. For large nuclei those charges and fields can become large enough to be treated classically.
However as two discs with charges move away from each other, those classical field are getting smaller and (in a still
poorly understood process) rather quickly create the quark-gluon plasma, in which the occupation
numbers are thermal, $O(1)$, with the classical fields disappearing. 
 
 Transition from field to plasma has been subject of multiple works, too many to be mentioned. One notable idea is existence of
instabilities \cite{Mrowczynski:1993qm}
such as the so called Weibel instability, producing filamentation of the longitudinal flow and
 transverse magnetic field $ B_\perp$. Asakawa et al \cite{Asakawa:2007gf}
argued that such chaotic magnetic field would remain in plasma, leading to
``abnormal viscosity". Unfortunately 
 those ideas, derived perturbatively and basically taken from electromagnetic plasma
contexts,  cannot possibly be valid for QGP at RHIC. 
 
 \begin{figure}[t]
  \begin{minipage}{5.cm}
\includegraphics[width=5cm]{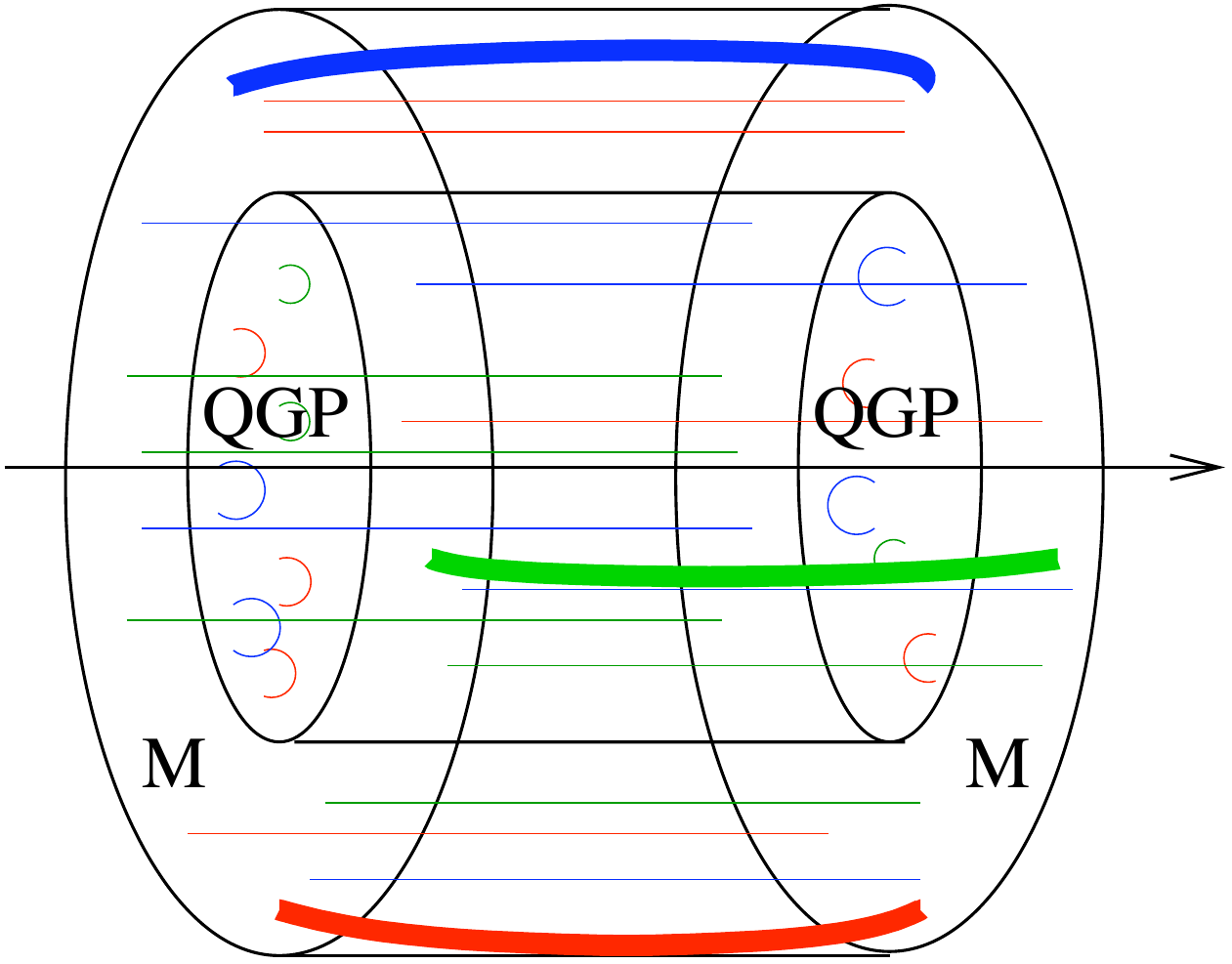}
\includegraphics[width=5cm]{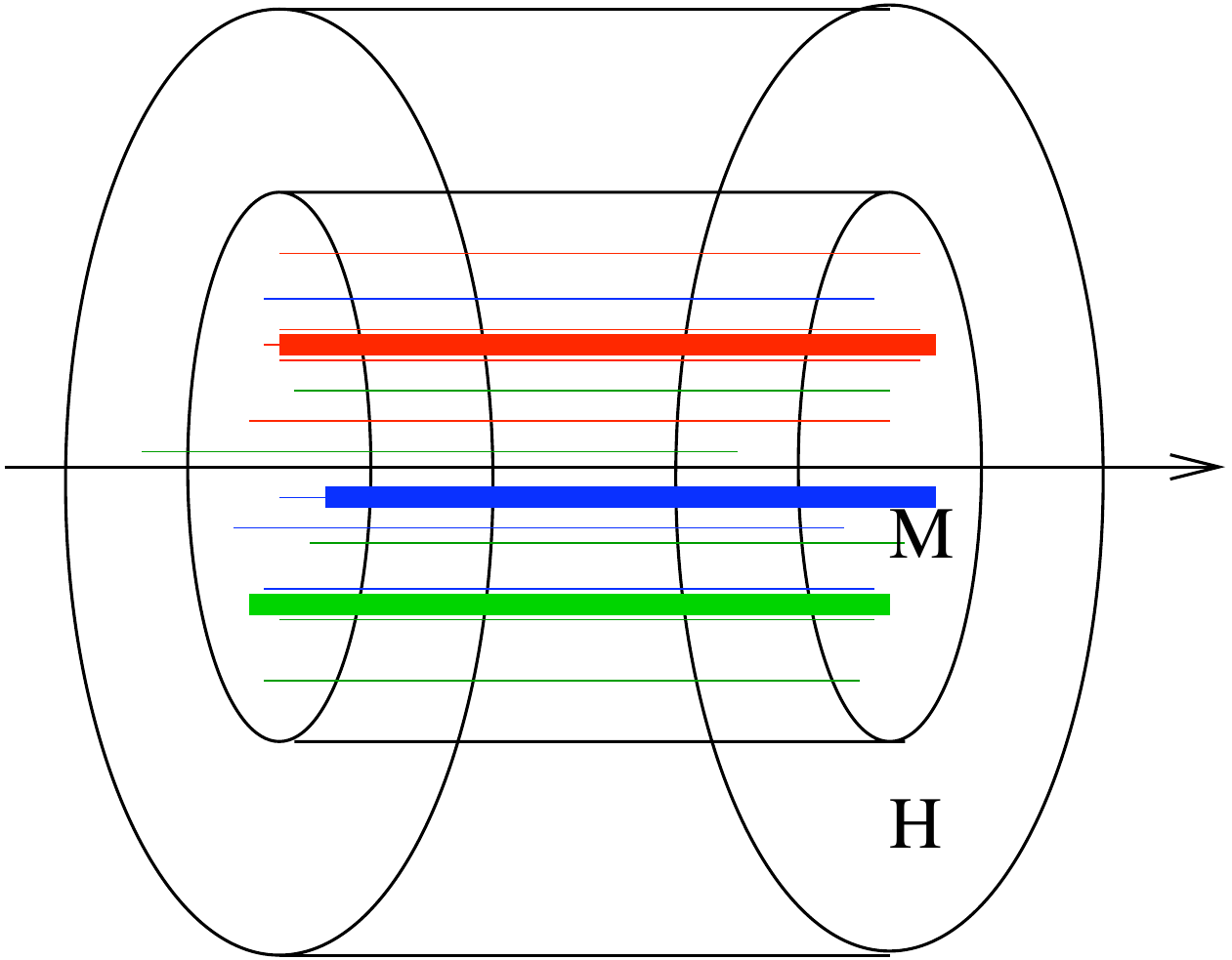}
\end{minipage}   \vskip -8.cm
\begin{minipage}{3.cm} \hspace*{7.cm}
\includegraphics[width=2.5cm]{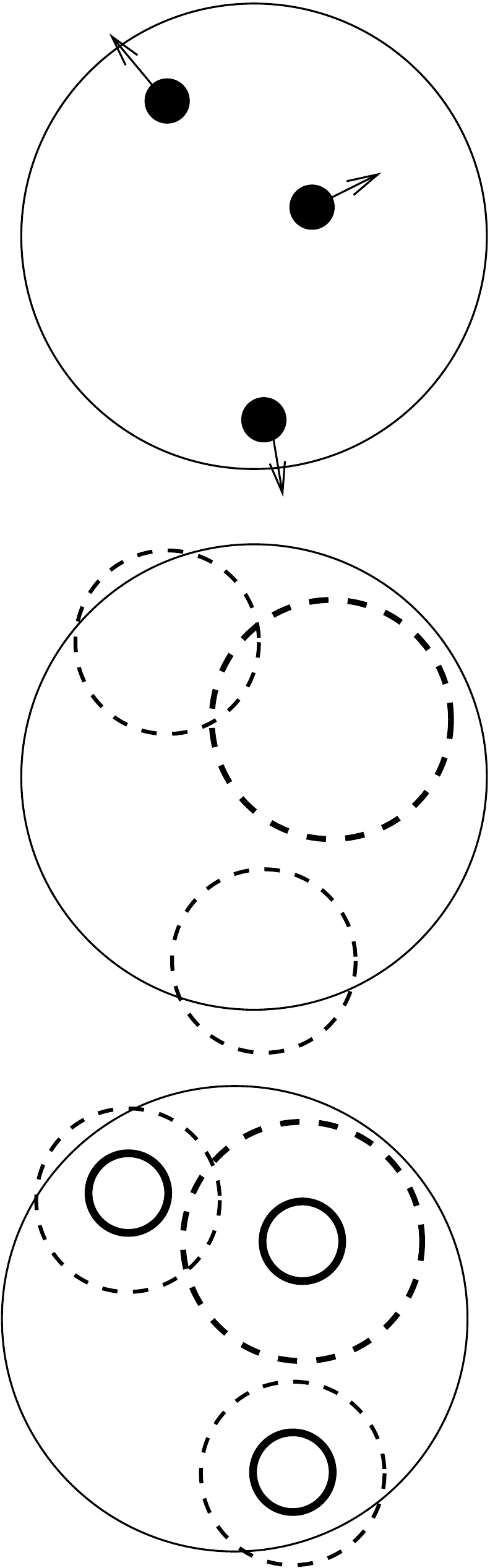}
\end{minipage}
\caption{(left)A snapshot of unscreened electric (dual-magnetic)
field in the M (near-$T_c$) region of the fireball. Upper and lower ones correspond to
full RHIC energy and the reduced energy (analogous to SPS).
(right)A sketch of the transverse plane of the colliding system:
the ``spots'' of extra density (a) are shown as black disks, to be moved
by collective radial flow (arrows). Naive sound expansion (b)
 would produce large-size and small amplitude wave: yet
the correct solution includes also brighter $secondary$ wave (c)
of smaller radius.}
\label{fig_dual_mag}
\end{figure}
 
  The crucial difference between the QED and QCD plasmas
 lies in the existence of magnetically charged quasiparticles -- monopoles and dyons --
leading to nonzero magnetic screening mass $M_M\sim g^2T$ first suggested by Polyakov long ago \cite{Polyakov:1978vu}
and by now well confirmed by subsequent lattice studies.
Thus hot QGP, unlike the electromagnetic plasmas,  screen $both$ the electric and the magnetic fields at some microscopic scales, although at a bit 
different ones.   
As at RHIC the central part of the produced fireball
reaches relatively high temperature $T\sim 2T_c$, we expect both $E,B$ fields to be effectively screened there,
see the central cylindrical part of Fig.\ref{fig_dual_mag}(upper left)  marked QGP.
But in the near-$T_c$ 
region  %vanishing electric component leads to vanishing electric screening mass. 
there is magnetic plasma in the outer  
cylindrical part of Fig.\ref{fig_dual_mag}  marked M (mixed or magnetic).% is nearly pure magnetic.
It is very important to emphasize that although this region on the phase diagram
is represented by a very narrow strip  $|T-T_c|\ll T_c$, it corresponds to more than order of variation
 of the energy or entropy density, and the corresponding space-time volume
 in the expansion of the fireball is by no means small. 
Another  snapshot of the geometry of the M region is shown in
 Fig.\ref{fig_dual_mag}(lower left): it corresponds to lower collision energy for which the M-phase resign inside the fireball:
 obviously this case is very different, it is planned to be investigated in a specialized RHIC run in 2011.

%%%%%%%%%%%%%
%recent theory developments known as ``magnetic scenario" \cite{Liao_ES_mono,Chernodub:2006gu} for the near-$T_c$
 % region show that
  %The situation with electric and magnetic screening masses get in fact inverted in this region.
   Multiple lattice studies, e.g.
ref \cite{Nakamura:2003pu} %(see also  Fig.16 my review \cite{Shuryak:2008eq}, where these issues are discussed in more detail)
 have shown that at  $T<1.4T_c$ the relation between electric and magnetic screening masses get inverted, namely $M_M>M_E$.
 As $T\rightarrow T_c$  the
       electric screening mass strongly decreases, partly because of heavy quark and gluon quasiparticles and partly because
       of their suppression by small  Polyakov loop expectation value $<L>$. So, near $T_c$
% the magnetic screening mass is  instead increasing monotonously, to a rather large value
 $ M_M(T\rightarrow T_c)\approx 3T_c,  M_E(T\rightarrow T_c)\approx 0 $.      
%Such behavior of  screening as well as other theoretical considerations had lead to the ``magnetic scenario" \cite{Liao_ES_mono,Chernodub:2006gu},
%which basically views the near-$T_c$ region, from about $0.8T_c$ up to 1.4$T_c$ as magnetic plasma, dominated by (gluomagnetic) monopoles. 
%The near-$T_c$ region, formally known as the ``M-phase" from ``mixed" of the 1st order transition, can thus still be called  ``the M-phase", now from ``magnetic". 
%While the
%QGP at $T>1.4T_c$ is in fact dominated by quarks and gluons,  it has been shown in \cite{Liao_ES_mono,Ratti:2008jz} (in classical and quantum settings respectively)
%that in a mixture of both electric and magnetic plasmas  the  electric-magnetic scattering dominates transport cross sections and viscosity till very high $T$,
%thus providing the natural explanation for unusually small
%viscosity observed at RHIC and predicting its value at the LHC. Its main reason is  the so called ``caging" phenomenon, induced by the Lorentz force.
The most important consequence  is that in the M-phase of the collision there is dense
 magnetic plasma, in which magnetic field is well screened while  $electric$ one remains $unscreened$ for a significant time.
This leads to the central idea: that QGP 
%produced in central part of RHIC collisions
 should have a ``dual corona" in which $electric$
(rather than magnetic) fields coexist with the plasma, affecting both the overall expansion and the propagation of perturbations.  
 We will suggest to use ``dual magnetohydrodynamics" (DMHD) for the description of diffuse electric fields in the M-phase, in particular
 study their effect
on the velocity of propagation of small perturbations. We will further argue below that like solar corona, that of QGP  should have metastable  flux tubes, although 
 microscopically thin ones which cannot be directly described by DMHD approximation. 

%%%%%%%
  Magnetohydrodynamics is a well known part of plasma physics,
developed by Alfven, Fermi, Chandrasekhar
and many others.
 It is an approximation which keeps
only one (magnetic) field nonzero and half of Maxwell eqns, while the other (electric) field is assumed to be totally screened. Ideal MHD approximation is 
 the limit of  $infinite$ conductivity of plasma
$\sigma\rightarrow\infty$, similar to $zero$ viscosity approximation for ideal hydrodynamics. In MHD the coupling between the field and 
and matter is obtained by inclusion of the
  (magnetic) field contribution into the stress tensor of the medium.
  
  Due to space limitations I will not put a complete set of resulting equations here, jumping directly to the main consequences. In a ``(dual)magnetized" plasma small amplitude perturbations are split
  into two modes, known as Alfven waves, propagating with two different speeds
\be u^2_{\pm}=({1\over 2}) \left[  {\vec{E}^2 \over  \epsilon} + c_s^2 \right]  
 \pm   ({1\over 2})   \left[    ({\vec{E}^2 \over  \epsilon} + c_s^2)^2 -  
 {4\vec{E}^2 cos^2\theta c_s^2   \over  \epsilon             } \right]^{1/2}            \ee
  where $E,c_s,\epsilon$ are the electric field, the  speed of sound in ``unmagnetized"  medium without field and plasma energy density, $\theta $ is the angle between the field strength  and the
the direction of the wave propagation. 
Note a case in which  the wave goes transverse to the
field (  $cos\theta=0$ ) in which the lower mode has $zero$ speed. 
 In the case of jet quenching -- when the jet direction is more or less up to 
experimentalist to pick --  general shape of these waves can be complicated. However when the jet (or original charge fluctuation)
propagates longitudinally, in the same direction as the field, the problem is axially symmetric and results in general
in two cones. The angles of their propagation can be obtained from the
previous expression, in which  the l.h.s. is substituted by Mach relation $u\rightarrow cos(\theta)v$, v is the velocity of the jet, and solve it for the $cos\theta$ (see the paper for more details on solutions).

If instead the electric field to be constant in space there is a a spot, localized in transverse
 plane, inevitable there is nonzero $\partial \vec{E}_z/\partial r$, which is a part of $curl(\vec{E})$ and by the (dual) Maxwell equation   
it is proportional to (dual, or magnetic) current $\tilde{j}_\phi$. 
This tells us that a flux tube solution must have a ``coil" with a current running around and  cancelling the field outside the spot. Ideal
DMHD has simple axially symmetric solution -- the macroscopic flux tube, with the field pressure balancing that of the plasma.
We dont discuss them in detail here because we think the flux tubes appeared in QGP corona
are actually {\em microscopically thin}, and so should be considered in a way worked out 
by Liao and myself previously \cite{Liao:2007mj,Liao:2008vj}.  Let me emphasize it again: we don't speak hear about well known flux tubes
at zero or low $T$ but about flux tubes which exist
at $T>T_c$, in the deconfined phase with magnetic constituents.  As the current is not a supercurrent in this case, they are not stable forever but are
metastable. We worked out quantum mechanics of the monopole-flux tube scattering and conditions for its mechanical stability.
Perhaps surprising to many, those flux tubes are $tighter$ (have larger tension) than the vacuum ones: this comes because uncondenced monopoles
have thermal momenta and their scattering creates larger pressure on the tube. As we will discuss at the end of the talk, those tubes in the M-phase
seem to have smaller breaking rate as well: so they are tighter in any sense. 

The original hints for their existence came from lattice data on finite-T potentials between static quark and antiquark, studied on the lattice.
 In brief, the central observation is large difference found between  the $free$ energy $F(T,r)$  and the $potential$ energy
$ V(T,r)=F(T,r)+TS(T,r) $
associated with quark pair at distance $r$.  While $F$ has no linear part (in $r$) above $T_c$, both $V$ and $S$ have it. 
The physical difference between the two
is that $F$ corresponds to adiabatically slow motion of the quarks, slow enough to produce maximal entropy possible and reach thermal
equilibrium at any $r$. However when quarks are moving with certain time scale, only a fraction $x$ of the
maximal entropy can be produced (because of Landau-Zener argument on level crossing): thus the effective potential is
$ V_{eff}(T,r)=F(T,r)+(1-x)TS(T,r) $. 
For relatively rapid motion in which no entropy is produced, $x=0$, and one returns to $V(T,r)$.
% These arguments have been  
%important to discussion of charmonium survival at RHIC as well as say dominance of baryons  \cite{our_suscept} in the near-$T_c$ region seen on the lattice.
The difference between the two potentials 
  can be also explained in other way\cite{Liao:2008vj}:  $F(T,r)$ is related to a stable flux tube with a ``supercurrent
coil''  while $V(T,r)$ has just ``normal"  coil  and thus is metastable . 

\section{QGP corona: experiment}

 Three different correlation phenomena have been discovered in
 heavy ion collisions at RHIC:\\

(i) The ``cone'' has been discovered in the 2-particle azimuthal
correlations . %like the one shown in Fig.\ref{fig_cone}.
% One can see from this figure
The ``away-side'' peak
 at $\Delta\phi=\pi$ due to a partner jet, balancing hard trigger particle, disappears, 
substituted by new peaks at completely
 different angle.%, as one moves from peripheral to central collisions.
After discovery of those effects there was extensive studies of the 3-particle
correlations, which %. This is  a rather complicated subject to go into here, let me just say that
%they has
 confirmed that the observed structure is indeed cone-like,  and not e.g. a reflected jet.

(ii) the hard ridge is also seen in 2-particle correlators,
but plotted on the two-dimensional $\Delta \phi-\Delta \eta$ plane,
the differences between the azimuthal angles and pseudorapidities
 of the two
particles.
The
 jet remnants make a 
peak  near   $\Delta \phi=0,\Delta \eta=0$, which was found to
sit on top of
the ``ridge'', with  comparable width in $\Delta \phi$ but
stretched in rapidity to nearly all 10 units (between beam rapidities at RHIC).
 For plots and various features one can consult
the original talk by  Putschke \cite{Putschke}. %Later it
%was shown by PHOBOS collaboration \cite{phobos} that the rapidity range of the ridge
%extends at least up to
%$|\eta|\approx 4$.

 (iii)  the ``soft ridge'' is found by STAR collaboration \cite{Adams:2005aw,star_ridge}
  without a trigger, in the 2-particle correlations.
For experimental details and phenomenological
considerations see% elated to these phenomena the reader may 
%consult e.g.
 %the
  talks at recent specialized workshop
 \cite{cathie_workshop}. %, see Fig.\ref{fig_peakwidth}.
%This happens because of two interrelated effects, 
%both well documented. For central collisions
%there is  (i) an $increase$ of the radial hydro velocity,
%accompanied by (ii)  a substantial $decrease$ in the freezeout
%temperature (which goes from  $T_c\approx 170 MeV$ in peripheral down to $T_f\approx 90 MeV$ for central
%collisions.

 We will  return to these observations  below,
 turning now to
 their suggested explanations:\\
(i) Stoecker et al,
 as well as Casalderrey, Teaney and myself \cite{conical} have proposed 
that the energy deposited by a quenched jet goes into two
hydrodynamical excitation modes, the sound and the so called diffusion or wake modes.   
The sound from the propagating jet  should thus  create the famous Mach cone,
in qualitative agreement  with the conical structure observed.  However surprising experimental fact is quite large amplitude of this signal, as well as large
 value of the cone anglle which has not been reproduced by hydrodynamics (or AdS/CFT).%.%, deduced from 2 and
 %3-particle correlators.
  The cone angle is %seems to be
 in the range $\theta_{cone}=1.2-1.4$ radians (not too far from $\pi/2=90^o$
or cylindrical waves!) which by
Mach formula  gives the  speed of pertinent perturbation to be about 
\be <v_{wave}>=cos(\theta_M)\approx 0.2 \ee
 while the expected speed of sound  is  .3 at its lowest point near $T_c$ and about .5 in QGP.
% We attribute it now to the ``second inner cone" due to DMHD effect. 
\\ 
(ii)  One early model for ``hard ridge'' has been 
introduced in my paper \cite{Shuryak:2007fu}.  It relates it with the
forward-backward jets accompanying any hard scattering,   providing
extra particles (``hot spot") widely distributed in rapidity.
This idea is then combined with the one  suggested previously by
Voloshin \cite{Voloshin:2003ud}, namely that extra particles
deposited in the fireball would be moved transversely by  the radial
hydrodynamical flow,  should produce a peak at certain
      azimuthal angle corresponding to the position of the hot spot,  see Fig.\ref{fig_dual_mag}(upper right).
While particles of the ridge are separated by large rapidity gaps and cannot
communicate during the expansion process,
their azimuthal emission angles remain correlated with each other because they originate from
 the same  ``hot spot'' in the transverse plane.\\
(iii) Similarly, transverse hydro boost of ``hot spots" was used for the
  explanation of the ``soft ridge''
  by McLerran and collaborators 
 \cite{Dumitru:2008wn,Gavin:2008ev}. They have pointed out that
the initial state color fluctuations in the
colliding nuclei would create  ``spots" at some positions in the transverse plane 
The idea is shown in Fig.\ref{fig_dual_mag}(upper right): the spots can be carried by hydro expansion and get visible at certain azimuthal angles. 
Further confirmation of hydro origin of ridges comes from the
centrality dependence of the {\em angular width} of the
ridge: the peak in azimuth $sharpens$ for
more central collisions.
% $without$ any hard collisions. 

So, at a very qualitative level the origin of all three phenomena seem to be explained:
yet at more qualitative level a lot of puzzles appear.  As an example, consider
the simplest of them, the ``soft ridge". As discussed in  \cite{Dumitru:2008wn,Gavin:2008ev},
 the initial stage   (proper time 
$\tau\sim 1/Q_s \sim 0.2\, fm/c$ where $Q_s\sim1\,GeV$ is the so called saturation scale
at RHIC) can be discussed using classical Yang-Mills
equations: thus color fluctuations naturally appear. 
However, the observed pions come from final freezeout time,
separated from the initial ``glasma" era by much longer time $\tau\sim 10\, fm$.
This is certainly so, as the explanation heavily relies on radial hydro velocity
and thus it has to wait till the hydro velocity is being created. As we will argue
below, there are many reasons why one might have expected nearly complete
disappearance of this signal during this time. 

Common to all three cases is deposition of 
 some additional energy (or entropy), on top of  the ``ambient matter". 
 The number of correlated particles in all of them 
constitute a 
small $(\sim 10^{-3})$ fraction of the total
multiplicity: thus they  can only be seen in a 
high-statistics correlation analysis. 

Similar to circles from
a stone thrown into a pond, initial
 perturbation should
 become some expanding waves, and hydrodynamics predicts that  basically nothing is left
  at the original location at later times.  Even ignoring any dissipation and using ideal hydrodynamics
  one finds 
the  radius of those waves at final freezeout to be given by
 the  ``sound horizon''
\be R_{h}=\int_0^{\tau_f} d\tau c_s(\tau)  \ee
So, by the  the freezeout proper time $\tau_f\sim 10-15\, fm/c$,
this distance is large, $\sim 6 fm$ or so, see Fig.\ref{fig_dual_mag}(right middle). 
% since the speed of sound changes between
%$c_s=1/\sqrt{3}\approx .58$ in QGP and about $.3$ at its minimum near $T_c$
 The amplitude of the sound wave is decreasing accordingly, and the width of $\phi$ distribution grows. 
  Simple calculation show that in such case the width of the peak would be larger than observed, and the amplitude smaller.
  This is a common puzzle for all three phenomena mentioned: hydro fails to explain why all of them remain observable,
%  t would not be possible to detect
%any trace of that tiny perturbation if it would be distributed over a significant fraction of the fireball:
%the only possibility is that it remains well localized in transverse direction.
%making us wandering if any trace of the perturbation can remain observable. 
%And yet, we $do$ observe all three correlations,
 as if nothing
happened to them during rather long time of the hydro process.
%$\sim 10 fm/c$. 
%This is the puzzle  discussion/resolution of which
i%s the main objective of this paper.  
The second set of waves provided by DMHD, with smaller velocity of propagation, or better yet stabilized flux tubes is thus suggested as an explanation, see Fig.\ref{fig_dual_mag}(bottom right) . 

Let me end this brief wandering into experiment-generated puzzles by mentioning
 recent RHIC data which provided direct experimental indications for enhanced stability of the flux tube in matter relative to pp. 
  Those  are from PHOBOS collaboration (now with a detector no longer physically existing)  which has large rapidity
coverage of their silicon detector.  They have studied clustering of secondaries in rapidity and found that the number of charged
particles in a cluster observed in AuAu collisions  is about twice that seen in correlation studies of the pp collisions, 
 reaching the size of about 10 particles. The rapidity width of the cluster is also larger than for those in pp: they are not
 consistent with isotropically decaying resonances but apparently elongated longitudinally. We basically see fragments of the 
 flux tubes, in pp and AuAu collisions: and apparently  
%(The exception are central collisions on the right, in which case extensive hadronic after-burning kills the correlations.) 
%The figure (b) shows that the produced clusters are not near-isotropically decaying resonances as in pp, (shaded horizontal region), but
%are instead more extended in rapidity. This last fact shows their
%direct relation to  the ``soft ridge" and flux tubes. Similar CuCu data (not shown) demonstrate basically the same clusters, provided
%the same centrality is taken: this shows that geometry and surface-to-volume ratio is important. 
 %Taken together, we interpret those clustering data as direct proof of significant changes in the flux tube decay parameters in AuAu relative to pp: 
the tubes  decay less frequently in the latter case, into larger pieces. I take it as the most direct indication to existence of the ``magnetic plasma" from the data.

\end{document}